\documentclass[twocolumn,a4paper,showpacs,prl,aps]{revtex4-1}

\usepackage[dvipsnames]{xcolor}
\usepackage{tikz}
\usetikzlibrary{chains,backgrounds,scopes,decorations.pathmorphing}
\usetikzlibrary{shapes.gates.logic.US,shapes}
\usetikzlibrary{arrows,patterns}
\usetikzlibrary{matrix}

 \usepackage{amsmath}
 \usepackage{amssymb}
 \usepackage{bbold}
 \usepackage{latexsym}
 \usepackage{amsfonts}
 \usepackage[caption=false]{subfig}
 \usepackage{epsfig}
 \usepackage{psfrag}
 \usepackage{color}
 \definecolor{darkblue}{rgb}{0,0,.5}
 \usepackage[linktocpage, colorlinks=true ,linkcolor=darkblue, citecolor=darkblue]{hyperref}
 \usepackage[all]{hypcap}

 \newcommand{\ket}[1]{\left|#1\right>}
 \newcommand{\bra}[1]{\left<#1\right|}
 \newcommand{\expval}[1]{\left< #1 \right>}
 
 \newcommand{\braket}[2]
 {\left<#1|#2\right>}
 \newcommand{\nn}{\nonumber\\}
 
 \newcommand{\f}[1]{\mbox{\boldmath$#1$}}

 \newcommand{\beq}{\begin{eqnarray}}
 \newcommand{\eeq}{\end{eqnarray}}
 \newcommand{\ord}{{\cal O}}
 \newcommand{\trace}[1]{{\rm Tr}\left\{ #1 \right\}}

 \newcommand{\abs}[1]{{\left| #1 \right|}}
 
 \newcommand{\HS}{\mathcal{H}_{\rm S}}
 \newcommand{\HB}{\mathcal{H}_{\rm B}}
 \newcommand{\HI}{\mathcal{H}_{\rm SB}}

 \newcommand{\dt}{\frac{\rm d}{\rm dt}}

 \newcommand{\ii}{{\rm i}}
\begin{document}

%\title{Signatures of Criticality in a non-equilibrium setup}
\title{Criticality in transport through the quantum Ising chain}

\author{Malte Vogl}\email{malte.vogl@tu-berlin.de}
\author{Gernot Schaller}\email{gernot.schaller@tu-berlin.de}
\author{Tobias Brandes}\email{tobias.brandes@tu-berlin.de}

\affiliation{Institut f\"ur Theoretische Physik, Technische Universit\"at Berlin, Hardenbergstr. 36, 10623 Berlin, Germany}

\begin{abstract}
We consider thermal transport between two reservoirs coupled by a quantum Ising chain as a model for non-equilibrium physics induced in quantum-critical many-body systems. 
By deriving rate equations based on exact expressions for the quasiparticle pairs generated during the transport, 
we observe signatures of the underlying quantum phase transition in the steady-state energy current already 
at finite and different reservoir temperatures. 
\end{abstract}

%05.30.Rt 	Quantum phase transitions (see also 64.70.Tg Quantum phase transitions in specific phase transitions; and 73.43.Nq Quantum phase transitions in Quantum Hall effects) 
%03.65.Yz 	Decoherence; open systems; quantum statistical methods (see also 03.67.Pp in quantum information; for decoherence in Bose-Einstein condensates, see 03.75.Gg)
%05.60.Gg 	Quantum transport 
%64.70.Tg 	Quantum phase transitions (for quantum Hall effects aspects, see 73.43.Nq in electronic structure of surfaces, interfaces, thin films, and low dimensional structures)


\maketitle
%%%%%%%%%%%%%%%%%%%%%%%%%%%%%%%%%%%%%%%%%%%%%%%%%%%%%%%%%%%%%%%%%%%%%%%%%%%%%%%%
Quantum phase transitions (QPTs) are drastic manifestations of quantum fluctuations in many-body systems that lead to 
critical behavior and states of matter with symmetry-broken phases~\cite{Sachdev}. 
Traditionally, such changes of state  are associated with changes in  ground state properties at zero temperature. 
Recently, considerable attention has turned to the role of excited states, as these become important 
if the fate of QPTs under modifications such as  finite temperature or non-equilibrium due to dissipation, 
external control, or driving is addressed. 
Among the fundamental issues then is the very definition of a QPT under non-equilibrium conditions, 
and furthermore the question of how non-analyticities of the ground state, phase diagrams, critical exponents, scaling behavior etc. are modified.

From the point of view of recent advances in simulations and realizations of QPTs in, e.g., cold atoms~\cite{Esslinger, Arimondo}, 
these are urgent issues, as experimental systems always contain a certain amount of dissipation and require external control 
and read-out mechanisms. 
For example, a classical external control allows one to explore novel states of matter and effective interactions which 
are absent in equilibrium~\cite{Lindner,Tanaka, Bastidas3, Zoller} and lead to new types of phase diagrams with 
additional phases, transition lines and critical points. 
Similar results have been obtained for QPTs in open, dissipative systems, either described by additional dissipative terms in equations 
of motions (similar to mean-field-type laser equations~\cite{MP2008,BMSK2012}) or Lindblad-type master equations, e.g.~\cite{Kessleretal2012,Hoening2012,DallaTorre2012}. %MV added citation
A further line of theoretical work concerning  excited-state phase transitions has also mostly been restricted to mean-field type QPTs 
(such as the Lipkin-Meshkov-Glick~\cite{PVM2008} or the  Dicke-superradiance~\cite{PeresFernandezetal2011} models), where  
non-analyticities in quantities like the density of states or excited state energies are essentially related to singularities in a (classical) energy landscape.
At zero system temperature, these can be made visible by suitable measurement devices -- as has e.g. been suggested for the Dicke model~\cite{lambert2009}.

In this letter, we investigate whether such features are also accessible in the non-equilibrium regime at the 
example of the quantum Ising chain in a transverse magnetic field.
Its interesting properties can either be studied in existing Ising ferromagnets~\cite{coldea2010a} or in 
corresponding quantum simulators, e.g.\ based on NMR techniques~\cite{zhang2009a}, trapped ions~\cite{friedenauer2008a}, atomic spins~\cite{edwards2010a,kim2011b}, or
electrons floating on liquid helium~\cite{mostame2008a}.
Quite some work has been devoted to thermal transport along open spin chains~\cite{wichterich2007a,prosen2008,prosen2010a,znidaric2011a,wu2011a}.
Complementing these approaches, we consider here a scenario where transport occurs perpendicularly through a closed ring of spins.
Going beyond linear response~\cite{mostame2007a}, this enables us to explore the extreme non-equilibrium regime.
We use the chain
% MV, which  has a transition from a paramagnetic into a $Z_2$ symmetry-broken ferromagnetic phase with long-range magnetic ordering at zero temperature, 
as a thermal coupling between two reservoirs whose temperature gradient drives a stationary, thermal current. 

Combining exact results for the Ising model with the  weak-coupling, Born-Markov-secular type description of 
open systems~\cite{breuerpetruccione2002}, 
%MV we demonstrate that the thermal current displays criticality at the same 
%value of the Ising chain coupling parameter as for the zero temperature (ground state) QPT\@.
%
% MV This has to be contrasted with stationary equilibrium quantities such as the average energy density or magnetization that display critical behavior only at $T=0$. 
%
% MV Traditionally, the vicinity of the critical point of equilibrium QPTs at finite temperatures has been characterized by spatio-temporal order parameter correlation functions, 
% which in particular for the quantum Ising chain are known to display a relatively complex behavior~\cite{Sachdev}. 
%GS moved
we find  that the stationary non-equilibrium current through the critical system itself may serve as a powerful tool to reveal 
finite-temperature quantum critical properties, that are not accessible in stationary system quantities such as the average energy density or magnetization. 
This does of course not exclude the possibility of much more complex features to be found in, e.g., 
temporal non-equilibrium correlations such as current noise spectra.
We expect these findings to generalize to other quantum critical systems which display a closing spectral gap.
% MV moved
We also find that the critical point for the phase transition is not changed by adding weak dissipation, 
provided the Lindblad operators are evaluated at the correct frequencies, i.e., in positivity-conserving secular approximation. 
%GS
Our conclusion from this observation is that phenomenological extensions of QPTs which simply add constant dissipation terms on top of equilibrium models have to be treated with caution.

%%%%%%%%%%%%%%%%%%%%%%%%%%%%%%%%%%%%%%%%%%%%%%%%%%%%%%%%%%%%%%%%%%%%%%%%%%%%%%%%%%%%%%%%%%%%%%%%%%%%%%%%%%%%%%%%%%%%%%%%%%%
%%%%%%%%%%%%%%%%%%%%%%%%%%%%%%%%%%%%%%%%%%%%%%%%%%%%%%%%%%%%%%%%%%%%%%%%%%%%%%%%%%%%%%%%%%%%%%%%%%%%%%%%%%%%%%%%%%%%%%%%%%%
%%%%%%%%%%%%%%%%%%%%%%%%%%%%%%%%%%%%%%%%%%%%%%%%%%%%%%%%%%%%%%%%%%%%%%%%%%%%%%%%%%%%%%%%%%%%%%%%%%%%%%%%%%%%%%%%%%%%%%%%%%%
%%%%%%%%%%%%%%%%%%%%%%%%%%%%%%%%%%%%%%%%%%%%%%%%%%%%%%%%%%%%%%%%%%%%%%%%%%%%%%%%%%%%%%%%%%%%%%%%%%%%%%%%%%%%%%%%%%%%%%%%%%%
{\em Model.---}
Our full model Hamiltonian $\mathcal{H} = \HS+\HI+\HB$ consists
of the quantum Ising chain in a transverse field for $N$ spins
\beq\label{eq:isingfull}
\HS=-g \sum_{i=1}^N \sigma_i^x - J \sum_{i=1}^N \sigma_i^z \sigma_{i+1}^z\,,
\eeq
where $g$ describes the coupling to an external magnetic field, $J$ the inter-chain coupling to nearest neighbors, and periodic boundary
conditions are assumed $\sigma^z_{N+1}\equiv\sigma^z_1$.
The model is analytically diagonalizable for finite $N$ and displays a second order quantum phase transition between a 
paramagnetic phase (for $g>J$) and a ferromagnetic one ($g<J$)~\cite{dziarmaga2005}.
Furthermore,  transport through the chain is enabled by coupling to two equilibrium reservoirs $\HB= \sum_{\alpha=S,D} \sum_q \omega_{\alpha q} b_{\alpha q}^\dagger b_{\alpha q}$, source (S) and drain (D), at different temperatures. Here, we consider non-interacting  bosons (e.g., photons or phonons) with momentum $q$ and energy $\omega_{\alpha q} $ (we set $\hbar=1$), having in mind a thermal-transport type flow of energy that is  mediated by  boson-induced spin-flips via the interaction
\beq\label{Hinter}
\HI = \sum_{\alpha,i,q} \sigma^x_i  \left[h_{\alpha q}^i b_{\alpha q}^\dagger + h_{\alpha q}^{i*} b_{\alpha q}\right]
\eeq
with interaction constants $h_{\alpha q}^i$ that depend on the specific realization of the model in an experimental context. 
%GS 
We mention that the  derivations below can also be carried out for fermionic reservoirs.
More importantly, we choose a particularly simple case of $\HI$  that is amenable for analytical calculation by assuming the coupling strengths 
as site-independent $h_{\alpha q}^i \to h_{\alpha q}$. 
Effectively, the coupling to the Ising chain is then via the  large spin operator $M_x \equiv \sum_i \sigma^x_i$.

\begin{figure}[t]
\includegraphics[width=0.45\textwidth,clip=true]{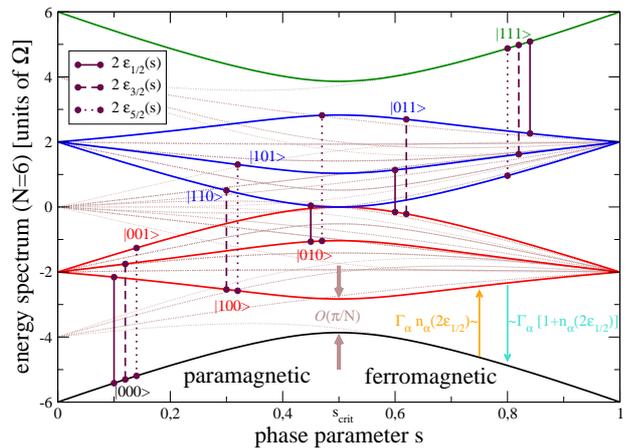}
\caption{(Color Online)
Spectrum of the quantum Ising chain of length $N=6$ versus phase parameter $s=J/\Omega=1-g/\Omega$ emphasizing 
the subspace of even quasi-particle numbers and vanishing pair momenta.
In the thermodynamic limit $N\to\infty$, the second order quantum phase transition comes with a vanishing energy gap above the ground state, cf. Eq.~(\ref{eq:energy}).
Possible transitions between states for second order perturbation in $h_{\alpha k}$ are marked by vertical lines with corresponding single-pair energies $2 \epsilon_k$.
Thin dotted curves in the background complete the full spectrum of Eq.~(\ref{eq:isingfull}).
\label{fig:sketch}
}
\end{figure}

We first diagonalize $\HS$ in the usual way, applying a Jordan-Wigner, Fourier, and Bogoliubov transformation (\cite{dziarmaga2005,appendix}) which, 
in the subspace of an even number of quasi-particles and for even $N$ (these %GS
constraints can in principle be lifted but will be tacitly assumed further-on),
transforms the system Hamiltonian into 
\mbox{$\HS' = \sum_k \epsilon_k (\gamma_k^\dagger \gamma_k - 1/2)$}
with fermionic annihilation operators $\gamma_k$~\cite{appendix}.
The quasi-momentum $k$ may assume half-integer values \mbox{$k=\pm 1/2, \pm 3/2, \ldots, \pm (N-1)/2$.}
The single-particle energies are defined by
\beq\label{eq:energy}
\epsilon_k = 2 \Omega \sqrt{(1-s)^2+s^2-2 s (1-s) \cos\left(\frac{2 \pi k}{N}\right)}\,,
\eeq
where we have introduced a dimensionless phase parameter by fixing $\Omega s=J$ and $\Omega(1-s)=g$ with energy scale $\Omega$, see Fig.~\ref{fig:sketch}.
%%%%%%%%%%%%%%%%%%%%%%%%%%%%%%%%%%%%%%%%%%%%%%%%%%%%%%%%%%%%%%%%%%%%%%%%%%%%%%%%
%%%%%%%%%%%%%%%%%%%%%%%%%%%%%%%%%%%%%%%%%%%%%%%%%%%%%%%%%%%%%%%%%%%%%%%%%%%%%%%%
%%%%%%%%%%%%%%%%%%%%%%%%%%%%%%%%%%%%%%%%%%%%%%%%%%%%%%%%%%%%%%%%%%%%%%%%%%%%%%%%
%%%%%%%%%%%%%%%%%%%%%%%%%%%%%%%%%%%%%%%%%%%%%%%%%%%%%%%%%%%%%%%%%%%%%%%%%%%%%%%%
%GS moved and modified
In the paramagnetic  phase with $s\ll s_{\rm crit}$, the quasiparticle pairs can be interpreted as spin waves running in opposite directions, 
whereas in the ferromagnetic domain $s\gg s_{\rm crit}$, they correspond to kinks between domains of opposite spin directions.

%GS paragraph about general 2 x 2 avoided crossings
QPTs are connected to a closing energy gap above the ground state, and a corresponding paradigmatic example is an avoided crossing 
between ground and first excited states for finite system sizes $N$ -- that becomes an exact crossing as $N\to\infty$.
Near the critical point $s_{\rm crit}$, the eigenvalues of ground state $\ket{0(s)}$ and first excited state $\ket{1(s)}$ can be expressed as 
$\lambda_0(s) = 1/2[\bar{\lambda}(s)-g(s)]$ and $\lambda_1(s)=1/2[\bar{\lambda}(s)+g(s)]$, where $\bar\lambda(s)$ is a linear (analytic) function of $s$ and 
the energy gap above the ground state $g(s)$ has a discontinuous first derivative as $N\to\infty$ at the critical point where also $g(s_{\rm crit}) = 0$.
Now, when coupling the quantum critical system to a single thermal reservoir by some operator $A$ at finite but sufficiently low temperature $\beta^{-1}$, the dynamics can be reduced 
to the subspace of ground and first excited state, and is described by a simple rate equation for the populations in the system energy eigenbasis
$\dot{P}_0(s) = -\gamma_{10}(s) P_0(s) + \gamma_{01}(s) P_1(s)$  and
$\dot{P}_1(s) = +\gamma_{10}(s) P_0(s) - \gamma_{01}(s) P_1(s)$, 
where the transition rates $\gamma_{ij}(s) = A(s) g_{ij}(s)$ depend on details of the coupling
$A(s) \equiv \abs{\bra{0(s)} A \ket{1(s)}}^2$ and the bath via $g_{ij}(s)$.
Importantly, we note that the stationary state solution does not depend on the matrix element $A(s)$.
%GS changed
Even more, provided the rate equation satisfies detailed balance $g_{10}(s)/g_{01}(s) = e^{-\beta g(s)}$, we see that due to the closing gap its
stationary solution is equipartitioned at the phase transition point in the continuum limit $\bar{P}_0(s_{\rm crit}) = \bar{P}_1(s_{\rm crit}) = 1/2$.
For e.g. the expectation value of the system energy $E(s) = \bar\lambda(s)/2 - g(s)/2 [P_0(s)-P_1(s)]$ we thus see that at finite temperatures 
the non-analytic properties of $g(s)$ will not be visible.
A similiar argument holds for general system operators with an analytic trace $\bar\lambda(s)$.
Thus, we generally do not expect local system observables to be good indicators for the ground state quantum-critical properties at finite temperatures.
In contrast, the current (that may be generated by coupling to two reservoirs at different temperatures) will be proportional to the matrix element
$A(s)$ and may thus inherit its critical dependence on $s$.
A toy model where these arguments are made explicit can be found in the supplement~\cite{appendix}.

Turning now to our model Hamiltonian, the very same transformations as used for the fermionic representation of $\HS$ map the interaction Hamiltonian, Eq.~(\ref{Hinter}), to~\cite{appendix}
\beq\label{eq:interaction}
%M_x' 
&& \HI' = \biggl\{ N \mathbb{1} - 2\sum_k \biggl[ |v_k|^2 \mathbb{1} +\left(|u_k|^2-|v_k|^2\right)\gamma_k^\dagger \gamma_k \\
&&+ u_k v_{-k} \left(\gamma_{+k}^\dagger \gamma_{-k}^\dagger +\gamma_{-k} \gamma_{+k} \right) \biggr]\biggr\}
\sum_{\alpha,q} \left[h_{\alpha q} b_{\alpha q}^\dagger + {\rm h.c.}\right] \nonumber
\,,
\eeq
where the coefficients are defined by \mbox{$v_k \propto s \sin\left(\frac{2\pi k}{N}\right)$} and 
\mbox{$u_k\propto \left[1-s -s \cos\left(\frac{2\pi k}{N}\right)+\epsilon_k/(2\Omega)\right]$} with
normalization $\abs{u_k}^2+\abs{v_k}^2=1$.
Obviously, the interaction $ \HI'$ does not couple subspaces with different values of the total
momentum $P = \sum_k k \gamma_k^\dagger \gamma_k$, since it
either does not create particles at all [first line in Eq.~(\ref{eq:interaction})]
or creates or annihilates only quasiparticle pairs of
opposite momenta.

Assuming that the system is initially prepared in the subspace of
vanishing total momentum (e.g., in its ground state $\ket{0}$), it
now suffices to consider the subspace of pairs with opposite
quasi-momenta only.
In this subspace, the basis elements can be conveniently constructed from the ground
state via
\beq\label{eq:basis}
\ket{\f{n}} = \ket{n_\frac{1}{2}, n_\frac{3}{2}, \ldots, n_\frac{N-1}{2}}
\equiv \prod_{k>0} \left(\gamma_{+k}^\dagger \gamma_{-k}^\dagger \right)^{n_k} \ket{0}\,,
\eeq
where $n_k \in \{0,1\}$ denotes the occupation of a quasi-particle pair with momenta $(+k,-k)$ such 
that $\left(\gamma_k^\dagger \gamma_k + \gamma_{-k}^\dagger \gamma_{-k}\right) \ket{\f{n}} = 2 n_k \ket{\f{n}}$.

% MV shifted for longer motivation of results
%GS cut down to fit with initial discussion
When both reservoirs are held at sufficiently low temperatures such that 
%GS definitions moved 
their bosonic occupations $n_\alpha(\omega) \equiv \left[e^{\beta_\alpha \omega} - 1\right]^{-1}$ vanish at all higher transition energies
$n_\alpha(2 \epsilon_k)\approx 0$ for all $k\ge 3/2$, 
the system will relax to the subspace spanned by the ground state $\ket{0}\equiv\ket{0,0,\ldots,0}$ and the first excited state \mbox{$\ket{1}\equiv\ket{1,0,\ldots,0}$}.
All states of higher occupations relax via the annihilation of higher-momentum quasi-particle pairs towards this subspace, see Fig.~\ref{fig:sketch}.
Then, the dynamics in this effective $2 \times 2$ subspace is governed by a rate equation of the previously discussed type
with rates $\gamma_{10}(s) = A(s) \sum_\alpha \Gamma_\alpha(g(s)) n_\alpha(g(s))$ and 
$\gamma_{01}(s) = A(s) \sum_\alpha \Gamma_\alpha(g(s)) \left[ 1+ n_\alpha(g(s)) \right]$, where the
spectral coupling densities $\Gamma_\alpha(\omega) \equiv 2\pi \sum_q \abs{h_{\alpha q}}^2 \delta(\omega-\omega_{\alpha q})$ and
the bosonic occupations depend on the phase parameter through the energy gap $g(s) = 2 \epsilon_{1/2}(s)$.
The overall matrix element equates to~\cite{appendix}
$A(s) = \abs{\bra{0} M_x \ket{1}}^2 = \frac{s^2 \sin^2\left(\frac{\pi}{N}\right)}{1-2s(1-s)\left[1+\cos\left(\frac{\pi}{N}\right)\right]}$, 
and becomes non-analytic at $s_{\rm crit}=1/2$ as $N\to\infty$.
Unfortunately, the argument to confine to just ground and first excited state breaks down for the Ising model, as for large $N$ also higher excited states
closely approach the ground state, cf. Eq.~(\ref{eq:energy}).
Since the simple picture of an avoided crossing is not strictly valid, we derive a more elaborate description for the Ising model in the following.

%%%%%%%%%%%%%%%%%%%%%%%%%%%%%%%%%%%%%%%%%%%%%%%%%%%%%%%%%%%%%%%%%%%%%%%%%%%%%%%%%%%%%%%%%%%%%%%%%%%%%%%%%%%%%%%%%%%%%%%%%%%
%%%%%%%%%%%%%%%%%%%%%%%%%%%%%%%%%%%%%%%%%%%%%%%%%%%%%%%%%%%%%%%%%%%%%%%%%%%%%%%%%%%%%%%%%%%%%%%%%%%%%%%%%%%%%%%%%%%%%%%%%%%
%%%%%%%%%%%%%%%%%%%%%%%%%%%%%%%%%%%%%%%%%%%%%%%%%%%%%%%%%%%%%%%%%%%%%%%%%%%%%%%%%%%%%%%%%%%%%%%%%%%%%%%%%%%%%%%%%%%%%%%%%%%
%%%%%%%%%%%%%%%%%%%%%%%%%%%%%%%%%%%%%%%%%%%%%%%%%%%%%%%%%%%%%%%%%%%%%%%%%%%%%%%%%%%%%%%%%%%%%%%%%%%%%%%%%%%%%%%%%%%%%%%%%%%
{\em High-Dimensional Rate Equation.---}
Applying lowest order perturbation theory in the coupling strength and in the relevant subspace, Eq.~\eqref{eq:basis}, (employing Born, Markov, and secular approximations~\cite{breuerpetruccione2002} in the standard way appropriate for open systems) yields a rate equation
\begin{equation}\label{rate_equation}
\dot{\rho}_{\f{n}} = \sum_{\f{m}} \left(\sum_\alpha \gamma^\alpha_{\f{n m}}\right) \rho_{\f{m}}
\end{equation}
for populations of the
system energy eigenstates \mbox{$\rho_{\f{n}} \equiv \bra{\f{n}} \rho \ket{\f{n}}$,} where the transition 
rates $\gamma^\alpha_{\f{n m}}$ due to reservoir $\alpha$ admit only creation or annihilation
of single quasi-particle pairs, see vertical lines in Fig.~\ref{fig:sketch}.
Assuming thermal reservoir states, the transition rates ($\f{n}\neq \f{m}$) evaluate to~\cite{appendix} 
\mbox{$\gamma^{\alpha}_{\f{nm}}=\Gamma_{\alpha}(\Delta_{\f{mn}})\left[1+n_{\alpha}(\Delta_{\f{mn}})\right]\abs{\bra{\f{n}}M_x\ket{\f{m}}}^2$} with
energy differences $\Delta_{\f{mn}}\equiv E_{\f{m}}-E_{\f{n}}$ and system energies \mbox{$E_{\f{n}} = \sum_{k>0} \epsilon_k (2 n_k - 1)$}.
The diagonal values $\gamma^\alpha_{\f{nn}}$ follow from trace conservation.
% removed GS
%, spectral coupling densities
%$\Gamma_\alpha(\omega)\equiv 2\pi \sum_q \abs{h_{\alpha q}}^2 \delta(\omega-\omega_q)$, 
%and bosonic occupations $n_\alpha(\omega) \equiv \left[e^{\beta_\alpha \omega}-1\right]^{-1}$ 
%(diagonal values $\gamma^\alpha_{\f{nn}}$ follow from trace conservation).
%
%

Using Eq.~(\ref{rate_equation}) and the rates $\gamma^\alpha_{\f{n m}}$, we obtain an analytical result for the non-equilibrium steady state solution~\cite{appendix},
\beq\label{eq:statstate}
\bar{\rho}_{\f{n}}=\prod_{k>0} \frac{\left[\bar{n}(2 \epsilon_k) \right]^{n_k}\left[1+\bar{n}(2 \epsilon_k) \right]^{1-n_k}}{1+2 \bar{n}(2 \epsilon_k)}\,,
\eeq
which -- similar to Ref.~\cite{vogl2011} -- is completely governed by an effective average 
bosonic occupation $\bar{n}(\omega) \equiv \frac{\sum_\alpha \Gamma_\alpha(\omega) n_\alpha(\omega)}{\sum_\alpha \Gamma_\alpha(\omega)}$.
However, our system has more than one allowed transition frequency, which implies that 
the stationary state~(\ref{eq:statstate}) is non-thermal (i.e., cannot be described by a single effective temperature) as soon as the reservoir temperatures are different 
[$n_S(\omega) \neq n_D(\omega)$].
We note that this non-equilibrium steady state for an interacting model holds for weak system-reservoir coupling only 
-- opposed to results obtained for non-interacting models~\cite{dhar2012}.
Eq.~(\ref{eq:statstate}) enables us to calculate the stationary values of the energy, the magnetization, and the current both for finite
chain lengths and in the thermodynamic limit $N\to\infty$.

%%%%%%%%%%%%%%%%%%%%%%%%%%%%%%%%%%%%%%%%%%%%%%%%%%%%%%%%%%%%%%%%%%%%%%%%%%%%%%%%%%%%%%%%%%%%%%%%%%%%%%%%%%%%%%%%%%%%%%%%%%%
%%%%%%%%%%%%%%%%%%%%%%%%%%%%%%%%%%%%%%%%%%%%%%%%%%%%%%%%%%%%%%%%%%%%%%%%%%%%%%%%%%%%%%%%%%%%%%%%%%%%%%%%%%%%%%%%%%%%%%%%%%%
%%%%%%%%%%%%%%%%%%%%%%%%%%%%%%%%%%%%%%%%%%%%%%%%%%%%%%%%%%%%%%%%%%%%%%%%%%%%%%%%%%%%%%%%%%%%%%%%%%%%%%%%%%%%%%%%%%%%%%%%%%%
%%%%%%%%%%%%%%%%%%%%%%%%%%%%%%%%%%%%%%%%%%%%%%%%%%%%%%%%%%%%%%%%%%%%%%%%%%%%%%%%%%%%%%%%%%%%%%%%%%%%%%%%%%%%%%%%%%%%%%%%%%%
{\em Energy and Magnetization.---}
For the mean stationary energy, we find~\cite{appendix}
\beq\label{eq:statenergy}
\bar{E} = \sum_{k>0} \frac{-\epsilon_k}{1+2\bar{n}(2 \epsilon_k)} \stackrel{N\to\infty}{\to} 
N \int\limits_0^{1/2} \frac{-\epsilon(\kappa)}{1+2 \bar{n}(2 \epsilon(\kappa))} d\kappa\,,
\eeq
where we have introduced the continuum of system energies $\epsilon(\kappa) \equiv \epsilon_{(N \kappa)}$.
At strictly zero temperature, where $\bar{n}(2\epsilon(\kappa))=0$, the system settles to the ground state, and the energy density 
can be expressed by a complete elliptic integral $E/N \to -\frac{2\Omega}{\pi} {\cal E}_E(4s(1-s))$, with a divergent second 
derivative at $s_{\rm crit}=1/2$.
This divergence, which reflects the usual ground state QPT criticality of the Ising chain, is also predictable from analyzing the analytic structure of the integrand 
in~(\ref{eq:statenergy}) at zero temperature.
For finite temperature and also in non-equilibrium setups where $\bar{n}(2\epsilon(\kappa)) \neq 0$, the energy density 
%GS modified
remains analytic.

We find a similar behavior for the magnetization, which for large $N$ becomes~\cite{appendix} ($v(\kappa) \equiv v_{(N k)}$)
\beq
\expval{M_x} \to N \left[1 - 4 \int\limits_0^{1/2} \frac{\abs{v(\kappa)}^2+\bar{n}(2\epsilon(\kappa))}{1+2 \bar{n}(2\epsilon(\kappa))} d\kappa\right]\,.
\eeq
At zero temperature, the integral is similarly solved by normal elliptic integrals and those of the first kind, which display a divergence in the
first derivative of the magnetization density with respect to $s$.
However, at finite temperature the magnetization density remains analytic, 
which is most evident in the trivial high-temperature case where $\bar{n}(2\epsilon(\kappa))\to\infty$.

%%%%%%%%%%%%%%%%%%%%%%%%%%%%%%%%%%%%%%%%%%%%%%%%%%%%%%%%%%%%%%%%%%%%%%%%%%%%%%%%%%%%%%%%%%%%%%%%%%%%%%%%%%%%%%%%%%%%%%%%%%%
%%%%%%%%%%%%%%%%%%%%%%%%%%%%%%%%%%%%%%%%%%%%%%%%%%%%%%%%%%%%%%%%%%%%%%%%%%%%%%%%%%%%%%%%%%%%%%%%%%%%%%%%%%%%%%%%%%%%%%%%%%%
%%%%%%%%%%%%%%%%%%%%%%%%%%%%%%%%%%%%%%%%%%%%%%%%%%%%%%%%%%%%%%%%%%%%%%%%%%%%%%%%%%%%%%%%%%%%%%%%%%%%%%%%%%%%%%%%%%%%%%%%%%%
%%%%%%%%%%%%%%%%%%%%%%%%%%%%%%%%%%%%%%%%%%%%%%%%%%%%%%%%%%%%%%%%%%%%%%%%%%%%%%%%%%%%%%%%%%%%%%%%%%%%%%%%%%%%%%%%%%%%%%%%%%%

{\em Heat Current.---}
This changes drastically, however, when we consider the heat current through the Ising chain from one reservoir to the other.
Analysis of the transition rates (e.g., by introducing energy counting fields as in~\cite{simine2012}) yields our main result for the current of net 
emitted bosons at the drain,
\beq\label{eq:current}
I&=& \sum_{\f{n,m}} (E_{\f{m}}-E_{\f{n}}) \gamma_{\f{nm}}^D \bar{\rho}_{\f{m}} \\
&=& \sum_{k>0} \frac{2 \epsilon_k A_k^2 \Gamma_S(2 \epsilon_k)\Gamma_D(2 \epsilon_k) \left[n_S(2 \epsilon_k)-n_D(2 \epsilon_k)\right]}
{\Gamma_S(2 \epsilon_k)\left[1+2 n_S(2 \epsilon_k)\right]+\Gamma_D(2 \epsilon_k)\left[1+2 n_D(2 \epsilon_k)\right]}, \nonumber
\eeq
where the second line follows after a straightforward calculation by inserting the stationary state and explicitly evaluating the transition rates~\cite{appendix}.
Here, we have introduced $A_k \equiv 4 u_k v_k = \frac{4 s \Omega}{\epsilon_k} \sin\left(\frac{2\pi k}{N}\right)$.

\begin{figure}[t]
\includegraphics[width=.45\textwidth,clip=true]{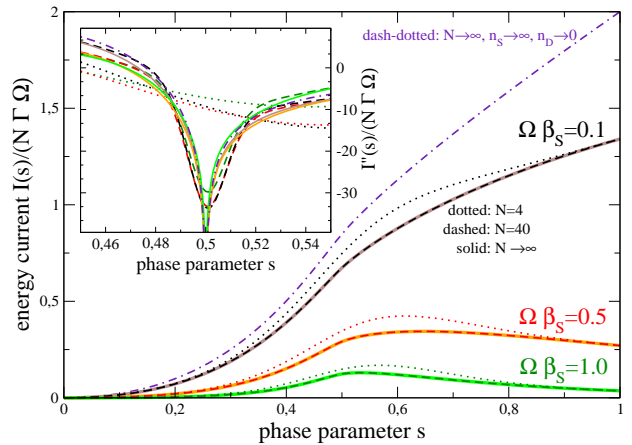}  
\caption{\label{fig:current}(Color Online)
Renormalized energy current $I$ and its second derivative w.r.t. $s$ (inset) versus phase parameter $s$ for different chain lengths $N=4,\,40,\,\infty$ (dotted, dashed, and bold solid, respectively) and
for different source temperatures $\Omega \beta_S =0.1, 0.5, 1.0$ (black/brown, red/orange, and dark/light green, respectively).
The dash-dotted purple curve denotes the analytically accessible case of $n_D(\omega) \to 0$, $n_S(\omega)\to\infty$, and
$N\to\infty$.
Other parameters: $\Omega\beta_D=10$,\,$\Gamma_S(\epsilon_k)=\Gamma_D(\epsilon_k)=\Gamma$.
}
\end{figure}

Evidently, the current is antisymmetric when $S\leftrightarrow D$, vanishes at equilibrium, and is positive when the source temperature exceeds the drain temperature 
[which implies $n_S(\omega) > n_D(\omega)$].
Most important however, in the thermodynamic limit $N\to \infty$ the current $I$ directly reflects the signatures of the ground state 
quantum phase transition of the Ising chain.
Formally, this correspondence is visible by the integral representation of $I$, which shows a divergence of its second derivative with respect to the phase parameter $s$ at all
temperatures, see Fig.~\ref{fig:current}.
The second derivative of the integrand in the continuum version $I/N\equiv\int_0^{1/2} j(s,\kappa) d\kappa$ of Eq.~(\ref{eq:current}) will at the critical point
$s_{\rm crit}=1/2$ for small $\kappa$ diverge as~\cite{appendix}
\beq
\left.\frac{\partial^2 j(s,\kappa)}{\partial s^2}\right|_{s=1/2} \approx -\frac{32 \Omega \Gamma_S \Gamma_D(\beta_D-\beta_S)}{\pi(\Gamma_S \beta_D+\Gamma_D \beta_S) \kappa}
+\ord\{\kappa\}\,,\qquad
\eeq
whilst the integrand itself and its first derivative remain finite.
Even for the extreme non-equilibrium, infinite thermobias regime the analytically obtainable~\cite{appendix} current
displays a divergence of the second derivative at $s_{\rm crit}=1/2$, compare the dash-dotted curves in Fig.~\ref{fig:current}.

{\em Conclusion.---} 
For transport through a closed Ising chain homogeneously coupled to two bosonic thermal reservoirs we have analytically 
shown that signatures of the underlying QPT are manifest in the energy current
already at finite temperature and also in the extreme non-equilibrium regime -- in contrast to system observables as mean energy or mean magnetization.
%GS moved here
We expect this result to generalize to a broader class of quantum critical systems, making the current a useful tool to analyze QPTs out of equilibrium. 
For finite sizes $N$, all quantities remain analytic but precursors of the QPT are already visible at rather moderate chain lengths.
Slightly perturbing the coupling symmetry, all subspaces of the model -- cf.\ thin curves in Fig.~\ref{fig:sketch} -- may be weakly coupled,
but since for near-multistable systems the separate current contributions are additive~\cite{schaller2010b}, 
we expect the sensitivity of the current to the underlying QPT to remain.
Furthermore, in the weak-coupling limit considered here, the position of the phase transition is neither changed nor are new phases introduced by the coupling to reservoirs.
Finally, we note that the subspace under consideration does not show a critical slow-down of relaxation, since every state is connected to the ground state by pair-annihilations only. In a coupling setup that does not preserve the parity, we expect the slow-down to occur.
%
% MV: Schon in Einleitung: We note that higher order transport cumulants (noise or time-dependent current autocorrelation functions) can be expected to yield 
% even more complex features and thus might prove useful tools to analyze QPTs out of equilibrium.

%%%%%%%%%%%%%%%%%%%%%%%%%%%%%%%%%%%%%%%%%%%%%%%%%%%%%%%%%%%%%%%%%%%%%%%%%%%%%%%%%%%%%%%%%%%%%%%%%%%%%%%%%%%%%%%%%%%%%%%%%%%
%%%%%%%%%%%%%%%%%%%%%%%%%%%%%%%%%%%%%%%%%%%%%%%%%%%%%%%%%%%%%%%%%%%%%%%%%%%%%%%%%%%%%%%%%%%%%%%%%%%%%%%%%%%%%%%%%%%%%%%%%%%
%%%%%%%%%%%%%%%%%%%%%%%%%%%%%%%%%%%%%%%%%%%%%%%%%%%%%%%%%%%%%%%%%%%%%%%%%%%%%%%%%%%%%%%%%%%%%%%%%%%%%%%%%%%%%%%%%%%%%%%%%%%
%%%%%%%%%%%%%%%%%%%%%%%%%%%%%%%%%%%%%%%%%%%%%%%%%%%%%%%%%%%%%%%%%%%%%%%%%%%%%%%%%%%%%%%%%%%%%%%%%%%%%%%%%%%%%%%%%%%%%%%%%%%

{\em Acknowledgments.---}
We acknowledge support by the DFG via GRK 1558 (M.V.), grants SCHA 1646/2-1 (G.S.), BRA 1528/7, BRA 1528/8, SFB 910 (T.B.).
The authors have benefited from discussions with V. Bastidas, M. Hayn, S. Kohler, G. Kie{\ss}lich, T. Novotn\'y and C. Nietner.

%%%%%%%%%%%%%%%%%%%%%%%%%%%%%%%%%%%%%%%%%%%%%%%%%%%%%%%%%%%%%%%%%%%%%%%%%%%%%%%%%%%%%%%%%%%%%%%%%%%%%%%%%%%%%%%%%%%%%%%%%%%
%%%%%%%%%%%%%%%%%%%%%%%%%%%%%%%%%%%%%%%%%%%%%%%%%%%%%%%%%%%%%%%%%%%%%%%%%%%%%%%%%%%%%%%%%%%%%%%%%%%%%%%%%%%%%%%%%%%%%%%%%%%
%%%%%%%%%%%%%%%%%%%%%%%%%%%%%%%%%%%%%%%%%%%%%%%%%%%%%%%%%%%%%%%%%%%%%%%%%%%%%%%%%%%%%%%%%%%%%%%%%%%%%%%%%%%%%%%%%%%%%%%%%%%
%%%%%%%%%%%%%%%%%%%%%%%%%%%%%%%%%%%%%%%%%%%%%%%%%%%%%%%%%%%%%%%%%%%%%%%%%%%%%%%%%%%%%%%%%%%%%%%%%%%%%%%%%%%%%%%%%%%%%%%%%%%

%%%%%%%%%%%%%%%%%%%%%%%%%%%%%%%%%%%%%%%%%%%%%%%%%%%%%%%%%%%%%%%%%%%%%%%%%%%%%%%%%%%%%%%%%%%%%%%%%%%%%%%%%%%%%%%%%%%%%%%%%%%
%%%%%%%%%%%%%%%%%%%%%%%%%%%%%%%%%%%%%%%%%%%%%%%%%%%%%%%%%%%%%%%%%%%%%%%%%%%%%%%%%%%%%%%%%%%%%%%%%%%%%%%%%%%%%%%%%%%%%%%%%%%
%%%%%%%%%%%%%%%%%%%%%%%%%%%%%%%%%%%%%%%%%%%%%%%%%%%%%%%%%%%%%%%%%%%%%%%%%%%%%%%%%%%%%%%%%%%%%%%%%%%%%%%%%%%%%%%%%%%%%%%%%%%
%%%%%%%%%%%%%%%%%%%%%%%%%%%%%%%%%%%%%%%%%%%%%%%%%%%%%%%%%%%%%%%%%%%%%%%%%%%%%%%%%%%%%%%%%%%%%%%%%%%%%%%%%%%%%%%%%%%%%%%%%%%

%%%%%%%%%%%%%%%%%%%%%%%%%%%%%%%%%%%%%%%%%%%%%%%%%%%%%%%%%%%%%%%%%%%%%%%%%%%%%%%%%%%%%%%%%%%%%%%%%%%%%%%%%%%%%%%%%%%%%%%%%%%
%%%%%%%%%%%%%%%%%%%%%%%%%%%%%%%%%%%%%%%%%%%%%%%%%%%%%%%%%%%%%%%%%%%%%%%%%%%%%%%%%%%%%%%%%%%%%%%%%%%%%%%%%%%%%%%%%%%%%%%%%%%
%%%%%%%%%%%%%%%%%%%%%%%%%%%%%%%%%%%%%%%%%%%%%%%%%%%%%%%%%%%%%%%%%%%%%%%%%%%%%%%%%%%%%%%%%%%%%%%%%%%%%%%%%%%%%%%%%%%%%%%%%%%
%%%%%%%%%%%%%%%%%%%%%%%%%%%%%%%%%%%%%%%%%%%%%%%%%%%%%%%%%%%%%%%%%%%%%%%%%%%%%%%%%%%%%%%%%%%%%%%%%%%%%%%%%%%%%%%%%%%%%%%%%%%

\appendix
\widetext
\newpage

\begin{center}
{Supplementary Material for\\
\Large ''Criticality in transport through the quantum Ising model''}\\
by M. Vogl, G. Schaller, and T. Brandes
\end{center}
\setcounter{equation}{0}
\setcounter{page}{1}

Equations in the main manuscript are referenced by square brackets.

%%%%%%%%%%%%%%%%%%%%%%%%%%%%%%%%%%%%%%%%%%%%%%%%%%%%%%%%%%%%%%%%%%%%%%%%%%%%%%%%%%%%%%%%%%%%%%%%%%%%%%%%%%%%%%%%%%%%%%%%%%%
%%%%%%%%%%%%%%%%%%%%%%%%%%%%%%%%%%%%%%%%%%%%%%%%%%%%%%%%%%%%%%%%%%%%%%%%%%%%%%%%%%%%%%%%%%%%%%%%%%%%%%%%%%%%%%%%%%%%%%%%%%%
\section{Introductory minimal model}

We consider a model for adiabatic quantum search [see Phys.\ Rev.\ A {\bf 65}, 042308 (2002)] as a minimal toy model that exhibits a 
first order quantum phase transition as the size $N$ of the system goes to infinity.
The Hamiltonian depends on the phase parameter $s$ and is given by
\beq
\HS &=& (1-s) \left[\mathbb{1} -\bigotimes_{i=1}^N \frac{\mathbb{1}_i + \sigma_i^x}{2} \right] 
+ s \left[\mathbb{1} - \bigotimes_{i=1}^N \frac{\mathbb{1}_i + (-1)^{w_i} \sigma_i^z}{2} \right]\nn
 &=& (1-s)\left[\mathbb{1} - \ket{p}\bra{p}\right]+ s \left[\mathbb{1} - \ket{w}\bra{w} \right]\,, 
\eeq 
where at $s=1$ the ground state $\ket{w}\equiv \ket{w_1,w_2,\dots,w_n}$ with $w_i \in \{0,1\}$ marks an arbitrary basis state in the so-called computational basis 
$\sigma^z_i \ket{w_i} = (-1)^{w_i} \ket{w_i}$.
For our purposes, it suffices to use e.g. $\ket{w}=\ket{0,\ldots,0}$.
In contrast, the ground state at $s=0$
\beq
\ket{p} = \frac{1}{\sqrt{2^N}}\sum_{z_1=0,1} \ldots \sum_{z_n=0,1} \ket{z_1,\ldots,z_n} = \bigotimes_{\ell=1}^N \frac{\ket{0}_\ell+\ket{1}_\ell}{\sqrt{2}}
\eeq
is the superposition of all basis states. 
We note that the overlap $\braket{p}{w}=\frac{1}{\sqrt{2^N}}$ is exponentially small in the system size. 
To represent $\HS$ in compact form, it is convenient to construct a suitable basis via 
Gram-Schmidt orthogonalization
\beq
\ket{\phi_0} &=& \ket{p}\,,\quad \ket{\phi_1} = \frac{\ket{w} - \braket{p}{w}\ket{p}}{\sqrt{1-|\braket{p}{w}|^2}} = \frac{\ket{w} - \frac{1}{\sqrt{2^N}}\ket{p}}{\sqrt{1-\frac{1}{2^N}}} = \ket{p_\perp}\,,
\eeq
and so on for all other states $\ket{\phi_2},\dots,\ket{\phi_{N-1}}$.
Importantly, the latter are also orthogonal to $\ket{w}$ and $\ket{p}$, such that the spectrum of $\HS$
is only nontrivial in the subspace spanned by $\ket{\phi_0}$ and $\ket{\phi_1}$.
In this subspace, the Hamiltonian reads
\beq
\HS' = 
\left(\begin{array}{cc}
s \left(1-2^{-N}\right) &  -s \sqrt{2^{-N} \left(1-2^{-N} \right)}\\
-s \sqrt{2^{-N} \left(1-2^{-N} \right)} & 1-s \left(1-2^{-N}\right)
\end{array}\right)\,,
\eeq
which allows to calculate the eigensystem $\HS'(s) \ket{\psi_\pm(s)} = \lambda_\pm(s) \ket{\psi_\pm(s)}$
\beq
\lambda_\pm (s) &=& \frac{1}{2} \left[1 \pm g(s) \right]\,,\quad 
\ket{\psi_\pm (s)}\propto \left[2^N-2 s \left(2^N-1\right) \pm 2^N g(s) \right] \ket{\phi_0} + 2 s \sqrt{2^N-1} \ket{\phi_1}\,,
\eeq
where $g(s) \equiv \sqrt{(1-2 s)^2+\frac{4}{2^N}s(1-s)}$ is the eigenvalue gap.
We see that when $N\to\infty$ the ground state energy changes its first derivative abruptly at the critical point $s_{\rm crit}=1/2$ (where
also the gap closes) and hence classify the quantum phase transition as first order [see also e.g. Quant.\ Inf.\ Comp.\ {\bf 10}, 0109 (2010)].

When in addition the system is coupled to bosonic reservoirs $\HB= \sum_{\alpha=S,D} \sum_q \omega_{\alpha q} b_{\alpha q}^\dagger b_{\alpha q}$ via 
\beq
\HI = \ket{p}\bra{p}\otimes \left[h_{\alpha q} b_{\alpha q}^\dagger + h_{\alpha q}^{*} b_{\alpha q}\right]\,,
\eeq
we see that the coupling does not leave the relevant subspace, such that with a compatible initial state 
we may neglect the states $\ket{\phi_2},\ldots$ in our considerations
even without invoking the low-temperature assumptions discussed in the main text.
We use standard techniques (applying Born-, Markov-, and secular approximations) to yield a rate equation for the diagonals of the system
density matrix in the system energy eigenbasis
\beq
\dt\left(\begin{array}{c}
\rho_{++}\\
\rho_{--}
\end{array}\right)
&=&
\abs{\bra{\psi_+(s)} \left(\ket{p}\bra{p}\right) \ket{\psi_-(s)}}^2 \sum_{\alpha=S,D} \Gamma_\alpha(s)
\left(\begin{array}{cc}
-n_\alpha(s) & 1+n_\alpha(s)\\
+ n_\alpha(s) & -1-n_\alpha(s)
\end{array}\right)
\left(\begin{array}{c}
\rho_{++}\\
\rho_{--}
\end{array}\right)\,,
\eeq
where the bare bosonic emission rates are given by $\Gamma_\alpha(s) \equiv 2\pi \sum_q \abs{h_{\alpha q}}^2 \delta(g(s)-\omega_{\alpha q})$, and 
$n_\alpha(s) = \left[e^{\beta_\alpha g(s)}-1\right]^{-1}$ denote the Bose-Einstein distribution of bath $\alpha$ evaluated at the energy gap.
The matrix element in front evaluates to
\beq
A(s)\equiv\abs{ \bra{\psi_+(g)} \left(\ket{p}\bra{p}\right) \ket{\psi_-(g)}}^2 = \frac{s^2 (2^N-1)}{2^{2N} g^2(s)}\,,
\eeq
which becomes non-analytic at $s_{\rm crit}=1/2$ as $N\to\infty$.
Clearly, the steady-state of this system does not depend on $A(s)$ 
\beq
\bar{\rho} = \left( \frac{\sum_\alpha \Gamma_\alpha(s) \left(1 + n_\alpha(s) \right)}{\sum_\alpha \Gamma_\alpha(s) \left(1 + 2 n_\alpha(s) \right)}, 
\frac{\sum_\alpha \Gamma_\alpha(s) n_\alpha(s)}{ \sum_\alpha \Gamma_\alpha(s) \left(1 + 2 n_\alpha (s) \right)}\right)^{\rm T}
\eeq
and is analytic when $\Gamma_\alpha(s)$ and $\Gamma_\alpha(s) n_\alpha(s)$ are analytic.
Using this result, we obtain for the stationary energy
\beq\label{Eegrover}
\bar{E}(s) = \expval{\HS'} = \sum_{i=+,-} \bra{\psi_i(s)} \HS' \ket{\psi_i(s)} \rho_{ii} 
= \frac{1}{2}\left(1 -g(s) \frac{\sum_\alpha \Gamma_\alpha(s)}{\sum_\alpha \Gamma_\alpha(s) (1+2 n_\alpha (s))}\right)\,.
\eeq
Similarly, we find for the expectation value of the system coupling operator
\beq\label{Eagrover}
\bar{A}(s) \equiv \expval{\ket{p}\bra{p}} =\frac{1}{2} \left(1+ \frac{\sum_\alpha \Gamma_\alpha(s)}{\sum_\alpha \Gamma_\alpha(s) (1+2 n_\alpha (s))} \frac{2^N (1-s)+2 s}{2^N g(s)} \right)\,.
\eeq
At strictly zero temperatures, where $n_\alpha(s)=0$, both expectation values (\ref{Eegrover}) and (\ref{Eagrover}) inherit the non-analytic properties of the energy gap $g(s)$.
For finite temperature, and assuming e.g. an Ohmic parametrization of the spectral coupling density (leading to finite transition rates)
$\Gamma_\alpha(s) = \Gamma_\alpha^0 g(s) e^{-g(s)/\omega_{\rm c}}$, we find for small energy gaps
\beq
\frac{\sum_\alpha \Gamma_\alpha(s)}{\sum_\alpha \Gamma_\alpha(s) (1+2 n_\alpha (s))} \stackrel{g(s)\to 0}{\to} 
\frac{\beta_S \beta_D (\Gamma_S^0+\Gamma_D^0)}{2(\beta_S \Gamma_D^0 + \beta_D \Gamma_S^0)} g(s) + \ord\{g^3(s)\}\,,
\eeq
which renders both expectation values analytic functions of $s$ near the critical point.
Generally, in the above expansion only odd powers of $g(s)$ occur, which means that both expectation values~(\ref{Eegrover}) and~(\ref{Eagrover})
can be expanded in powers of $g^2(s)$ and are thus -- due to the absence of the root -- analytic functions of $s$.

Using e.g. the formalism of Full Counting Statistics, the stationary bosonic current through the system into the drain reservoir can also be calculated
\beq
\bar{I}(s) = A(s) \frac{\Gamma_S(s) \Gamma_D(s) \left[n_S(s) -n_D(s)\right]}{\sum_\alpha \Gamma_\alpha(s) \left[1+2 n_\alpha(s)\right]}\,,
\eeq
and its first derivative becomes nonanalytic at the critical point for all non-equilibrium temperature configurations already due to the dependence of the
prefactor $A(s)$.
Eventually, the non-analyticity of the current can thus be directly related to the closing of the energy gap.

%%%%%%%%%%%%%%%%%%%%%%%%%%%%%%%%%%%%%%%%%%%%%%%%%%%%%%%%%%%%%%%%%%%%%%%%%%%%%%%%%%%%%%%%%%%%%%%%%%%%%%%%%%%%%%%%%%%%%%%%%%%
%%%%%%%%%%%%%%%%%%%%%%%%%%%%%%%%%%%%%%%%%%%%%%%%%%%%%%%%%%%%%%%%%%%%%%%%%%%%%%%%%%%%%%%%%%%%%%%%%%%%%%%%%%%%%%%%%%%%%%%%%%%
\section{The quantum Ising model}
For the quantum Ising model a two-level approximation is not appropriate for the case of thermal transport. In the following we will derive analytic results for the full-dimensional problem.

\subsection{Mapping the system Hamiltonian to non-interacting fermions}

The Jordan-Wigner transform (JWT)
\beq\label{eq:jwt}
\sigma^x_n = \f{1} - 2 c_n^\dagger c_n\,,\qquad
\sigma^z_n = -(c_n+c_n^\dagger) \prod_{m=1}^{n-1} \left(\f{1}-2 c_m^\dagger c_m\right)
\eeq
maps the spin-1/2 Pauli matrices non-locally to fermionic operators $c_m$.
Inserting the JWT into the Hamiltonian~[1], one has to treat the boundary term with special care
\beq\label{Ehamilton2}
H &=& -g \sum_{n=1}^N (\f{1}-2c_n^\dagger c_n) 
- J \sum_{n=1}^{N-1} (c_n+c_n^\dagger)(c_{n+1}+c_{n+1}^\dagger) (\f{1}-2 c_n^\dagger c_n)
- J (c_N+c_N^\dagger) \left[\prod_{n=1}^{N-1} (\f{1}-2 c_n^\dagger c_n)\right] (c_1+c_1^\dagger)\nn
&=& -g \sum_{n=1}^N (\f{1}-2c_n^\dagger c_n) 
- J \sum_{n=1}^{N-1} (c_n^\dagger-c_n)(c_{n+1}^\dagger+c_{n+1})
+ J (c_N^\dagger-c_N)(c_1^\dagger+c_1) \left[\prod_{n=1}^N (\f{1}-2 c_n^\dagger c_n)\right]\,,
\eeq
where we have extensively used the fermionic anticommutation relations.
Introducing the projection operators on the subspaces with even (+) and odd (-) total number
of fermion quasiparticles
\beq
{\cal P}^{\pm} \equiv \frac{1}{2}\left[\f{1}\pm\prod_{m=1}^N (\f{1}-2 c_m^\dagger c_m)\right]\,,
\eeq
we may also write the Hamiltonian (\ref{Ehamilton2}) 
$H = ({\cal P}^+ + {\cal P}^-) H  ({\cal P}^+ + {\cal P}^-)$.
It is straightforward to see that terms with different projectors and with $n<N$ vanish
\beq
0 &=& {\cal P}^+ (\f{1}-2 c_n^\dagger c_n) {\cal P}^- = {\cal P}^- (\f{1}-2 c_n^\dagger c_n) {\cal P}^+\,,\nn
0 &=& {\cal P}^+ (c_n^\dagger-c_n)(c_{n+1}^\dagger+c_{n+1}) {\cal P}^- = {\cal P}^- (c_n^\dagger-c_n)(c_{n+1}^\dagger+c_{n+1}) {\cal P}^+\,.
\eeq
For the boundary terms one finds similarly
\beq
&&({\cal P}^++{\cal P}^-) (c_N^\dagger-c_N) (c_1^\dagger + c_1)  \left[\prod_{n=1}^N (\f{1}-2 c_n^\dagger c_n)\right] ({\cal P}^++{\cal P}^-)\nn
&=&({\cal P}^++{\cal P}^-) (c_N^\dagger-c_N) (c_1^\dagger + c_1) (2 {\cal P}^+-\f{1}) ({\cal P}^++{\cal P}^-)\nn
&=&{\cal P}^+ (c_N^\dagger-c_N)(c_1^\dagger+c_1) {\cal P}^+ - {\cal P}^- (c_N^\dagger-c_N)(c_1^\dagger+c_1) {\cal P}^-\,,
\eeq
such that we can finally write the Hamiltonian (\ref{Ehamilton2}) as the sum of two non-interacting parts with either an even or an odd total 
number of fermionic quasiparticles
\beq\label{Ehamilton3}
H &=& {\cal P}^+ H^+ {\cal P}^+ +  {\cal P}^- H^- {\cal P}^-\nn
&=& {\cal P}^+ \left[-g \sum_{n=1}^N (\f{1}-2 c_n^\dagger c_n)-J \sum_{n=1}^N (c_n^\dagger - c_n)(c_{n+1}^\dagger+c_{n+1})\right] {\cal P}^+\nn
&&+ {\cal P}^- \left[-g \sum_{n=1}^N (\f{1}-2 c_n^\dagger c_n)-J \sum_{n=1}^N (c_n^\dagger - c_n)(c_{n+1}^\dagger+c_{n+1})\right] {\cal P}^-\,.
\eeq
Note that this requires to define antiperiodic boundary conditions in the even (+) subspace $c_{N+1}^{(+)}\equiv - c_1^{(+)}$ and periodic boundary conditions in the odd (-) 
subspace  $c_{N+1}^{(-)}\equiv + c_1^{(-)}$.

Since the even subspace is relevant to our model, we further seek to diagonalize the Hamiltonian
\beq\label{Ehamilton4}
H^+ = -g \sum_{n=1}^N (\f{1}-2 c_n^\dagger c_n) - J \sum_{n=1}^N (c_n^\dagger-c_n)(c_{n+1}^\dagger+c_{n+1})
\eeq
with antiperiodic boundary conditions $c_{N+1}=-c_1$.
Translational invariance suggests to use the discrete Fourier transform (DFT, preserving the anticommutation relations
due to its unitarity by construction)
\beq\label{eq:dft}
c_n = \frac{e^{-i \pi/4}}{\sqrt{N}} \sum_k \tilde{c}_k e^{+i k n \frac{2\pi}{N}}\,,
\eeq
which is compatible with the antiperiodic boundary conditions when $k$ takes half-integer values
\beq
k \in \{\pm \frac{1}{2}, \pm \frac{3}{2}, \pm \frac{5}{2}, \ldots\}\,,\qquad\mbox{where}\qquad\abs{k} \le \frac{N-1}{2}
\eeq
(Note that the number of quasiparticles in the even subspace is the same e.g.\ for $N=6$ and $N=7$).
The DFT maps the Hamiltonian into
\beq
H^+ &=& -g N \f{1} + \sum_k \left\{2 [g-J \cos(k 2 \pi/N) ] \tilde{c}_k^\dagger \tilde{c}_k + J \sin(k 2\pi/N) \left[\tilde{c}_k^\dagger \tilde{c}_{-k}^\dagger + \tilde{c}_{-k} \tilde{c}_k\right]\right\}\,.
\eeq
Now, the observation that only positive and negative frequencies couple (conservation of one-dimensional quasi-momentum), suggests to use the reduced Bogoliubov transform
\beq\label{eq:bogoliubov}
\tilde{c}_k = u_{+k} \gamma_{+k} + v_{-k}^* \gamma_{-k}^\dagger\,,
\eeq
which mixes positive and negative momenta and where the a priori unknown coefficients have already been labeled suggestively (a more general ansatz would eventually
of course yield the same solution).
Since the new operators $\gamma_k$ should be fermionic, we obtain from the orthonormality conditions
\beq
1 = \abs{u_{+k}}^2 + \abs{v_{-k}}^2\,,\qquad
0 = u_{+k} v_{+k}^* + u_{-k} v_{-k}^*
=
(v_{+k}^*,v_{-k}^*)
\left(\begin{array}{c}
u_{+k}\\
u_{-k}
\end{array}\right)
 \,.
\eeq
Demanding that the Bogoliubov transform eliminates all non-diagonal terms (of the form $\gamma_{-k} \gamma_{+k}$ etc.) yields (by combining positive and negative $k$) the
equation
\beq
0 &=& 2 \left[g - J \cos\left(k \frac{2\pi}{N}\right)\right] \left(u_{+k} v_{-k} - u_{-k} v_{+k}\right) 
+ 2 J \sin\left(k \frac{2\pi}{N}\right) \left(u_{-k} u_{+k} + v_{-k} v_{+k}\right)\nn
&=& \left(v_{-k},u_{-k}\right)
\left(\begin{array}{cc}
+ 2 \left[g - J \cos\left(k \frac{2\pi}{N}\right)\right] & + 2 J \sin\left(k \frac{2\pi}{N}\right)\\
+ 2 J \sin\left(k \frac{2\pi}{N}\right) & -  2 \left[g - J \cos\left(k \frac{2\pi}{N}\right)\right]
\end{array}\right)
\left(\begin{array}{c}
u_{+k}\\
v_{+k}
\end{array}\right)
\equiv
\left(v_{-k},u_{-k}\right)
{\cal M}
\left(\begin{array}{c}
u_{+k}\\
v_{+k}
\end{array}\right)
\,.
\eeq
All equations can be fulfilled when we choose $(u_{+k},v_{+k})^T$ as the normalized positive energy eigenstate of the matrix ${\cal M}$ with eigenvalue
\beq\label{Espeigenvalue}
\epsilon_k^+ = +2 \sqrt{g^2+J^2-2 g J \cos(k 2\pi/N)} \equiv \epsilon_k
\eeq
and
$(v_{-k}^*,u_{-k}^*)^T=(-v_{+k}^*,+u_{+k}^*)^T$ as its negative energy eigenstate with eigenvalue $\epsilon_k^- = -2 \sqrt{g^2+J^2-2 g J \cos(k 2\pi/N)}$.
To be more explicit, we have
\beq\label{eq:coefficients}
u_k &=& \frac{g - J \cos(k 2\pi/N) +\sqrt{g^2+J^2-2 g J \cos(k 2\pi/N)}}{\sqrt{\left[g - J \cos(k 2\pi/N) +\sqrt{g^2+J^2-2 g J \cos(k 2\pi/N)}\right]^2+\left[J \sin(k 2\pi/N)\right]^2}
}\,,\nn
v_k &=& \frac{J \sin(k 2\pi/N)}{\sqrt{\left[g - J \cos(k 2\pi/N) +\sqrt{g^2+J^2-2 g J \cos(k 2\pi/N)}\right]^2+\left[J \sin(k 2\pi/N)\right]^2}}\,.
\eeq
Using these solutions, we obtain when $N$ is even
\beq\label{Ehamplus}
H^+ &=& \sum_k 2 \sqrt{g^2+J^2-2 g J \cos\left(k \frac{2\pi}{N}\right)} \left(\gamma_k^\dagger \gamma_k - \frac{1}{2}\right) = \sum_k \epsilon_k \left(\gamma_k^\dagger \gamma_k - \frac{1}{2}\right)\,,
\eeq
which reproduces the inline equation before~[3] in the main manuscript.

%%%%%%%%%%%%%%%%%%%%%%%%%%%%%%%%%%%%%%%%%%%%%%%%%%%%%%%%%%%%%%%%%%%%%%%%%%%%%%%%%%%%%%%%%%%%%%%%%%%%%%%%%%%%%%%%%%%%%%%%%%%
%%%%%%%%%%%%%%%%%%%%%%%%%%%%%%%%%%%%%%%%%%%%%%%%%%%%%%%%%%%%%%%%%%%%%%%%%%%%%%%%%%%%%%%%%%%%%%%%%%%%%%%%%%%%%%%%%%%%%%%%%%%

\subsection{Mapping of the interaction Hamiltonian}

Obviously, the used transformations do not affect the reservoir part, such that it suffices to transform $M_x = \sum_{i=1}^N \sigma^x_i$
with the very same transformations as before.
Inserting the JWT~(\ref{eq:jwt}) yields
\beq
M_x = N \f{1} - 2 \sum_{n=1}^N c_n^\dagger c_n\,.
\eeq
Furthermore, inserting the DFT~(\ref{eq:dft}) leads to
\beq
M_x = N \f{1} - 2 \sum_k \tilde{c}_k^\dagger \tilde{c}_k\,.
\eeq
Finally, inserting the Bogoliubov transformation~(\ref{eq:bogoliubov}), replacing $k\to-k$ in some terms and exploiting that the coefficients~(\ref{eq:coefficients}) are real
yields
\beq
M_x = N \f{1} - 2 \sum_k \left[\abs{u_k}^2 \gamma_k^\dagger \gamma_k + \abs{v_k}^2 \gamma_k \gamma_k^\dagger 
+ u_k v_{-k} \left(\gamma_{+k}^\dagger \gamma_{-k}^\dagger + \gamma_{-k} \gamma_{+k}\right)\right]\,,
\eeq
which by using the fermionic anticommutation relations is equivalent to~[4] in the main manuscript.
%
%CORRECT SIGN ERROR!! ->Corrected : Malte 17.8.
%
For later convenience we write this as a sum over positive $k$-values only
\beq\label{eq:mxreduced}
M_x = N \f{1} - 4 \sum_{k>0} \abs{v_k}^2 \f{1} - 2 \sum_{k>0} \left(1-2\abs{v_k}^2\right) \left(\gamma_k^\dagger \gamma_k + \gamma_{-k}^\dagger \gamma_{-k}\right)
+ 4 \sum_{k>0} u_k v_k \left(-\gamma_{+k}^\dagger \gamma_{-k}^\dagger + \gamma_{+k}\gamma_{-k}\right)\,.
\eeq

%%%%%%%%%%%%%%%%%%%%%%%%%%%%%%%%%%%%%%%%%%%%%%%%%%%%%%%%%%%%%%%%%%%%%%%%%%%%%%%%%%%%%%%%%%%%%%%%%%%%%%%%%%%%%%%%%%%%%%%%%%%
%%%%%%%%%%%%%%%%%%%%%%%%%%%%%%%%%%%%%%%%%%%%%%%%%%%%%%%%%%%%%%%%%%%%%%%%%%%%%%%%%%%%%%%%%%%%%%%%%%%%%%%%%%%%%%%%%%%%%%%%%%%

\subsection{Derivation of the rate equation}

We rely on previous results in the literature that yield for an interaction Hamiltonian of the form
$\HI= A\otimes B$ under Born, Markov, and secular approximations a Lindblad-type master equation.
When the spectrum of the system Hamiltonian is non-degenerate (and here more specifically, 
when states coupled in the master equation are energetically non-degenerate), this Lindblad master 
equation couples only the diagonal elements of the density matrix in the system energy eigenbasis to each other, i.e., 
it can be written in the form of a rate equation
\beq
\dot{\rho}_{aa} = \sum_b \gamma_{ab,ab} \rho_{bb} - \sum_b \gamma_{ba,ba} \rho_{aa}\,,
\eeq
where $a,b$ label the different system energy eigenstates.
Note that the refined condition of non-degeneracy is for finite $N$ always fulfilled, as e.g.\ the intersection point in Fig.~1 between $\ket{001}$ and $\ket{110}$ 
in the main manuscript are between uncoupled states.
The transition rates are given by
\beq
\gamma_{ab,ab} = \gamma(E_b-E_a) \abs{\bra{a} A \ket{b}}^2\,,
\eeq
where $\gamma(\omega)$ is the Fourier transform of the bath correlation function,
$\gamma(\omega) \equiv \int C(\tau) e^{+\ii\omega\tau} d\tau \equiv \int \expval{e^{+\ii \HS \tau} B e^{-\ii \HS \tau} B} e^{+\ii\omega\tau} d\tau$.
Specifically for our model there is only one, which for two thermal source and drain reservoirs becomes
\beq
C(\tau) &=& \sum_{\alpha\alpha' qq'} \expval{\left[h_{\alpha q} b_{\alpha q}^\dagger e^{+\ii \omega_{\alpha q} \tau} + {\rm h.c.}\right]
\left[h_{\alpha' q'} b_{\alpha' q'}^\dagger + {\rm h.c.}\right]}
= \sum_\alpha \sum_q  \abs{h_{\alpha q}}^2 \left[\expval{b_{\alpha q}^\dagger b_{\alpha q}} e^{+\ii \omega_{\alpha q} \tau} 
+ \expval{b_{\alpha q} b_{\alpha q}^\dagger} e^{-\ii \omega_{\alpha q} \tau}\right]\nn
&=& \frac{1}{2\pi} \sum_\alpha \int\limits_0^\infty \Gamma_\alpha(\omega) \left[n_\alpha(\omega) e^{+\ii\omega\tau} + (1+n_\alpha(\omega)) e^{-\ii\omega\tau}\right] d\omega\,,
\eeq
where we have introduced for $\omega>0$ the spectral coupling density $\Gamma_\alpha(\omega) \equiv 2\pi \sum_q \abs{h_{\alpha q}}^2 \delta(\omega - \omega_{\alpha q})$ (recall
that $\omega_{\alpha q}>0$) and $n_\alpha(\omega) \equiv \left[e^{\beta_\alpha\omega} -1\right]^{-1}$ denotes the Bose distribution for reservoir $\alpha$ held at inverse temperature $\beta_\alpha$.
Analytically continuing the spectral coupling densities to negative frequencies via $\Gamma_\alpha(-\omega) \equiv - \Gamma_\alpha(+\omega)$ 
and exploiting that $n_\alpha(-\omega) = -(1+n_\alpha(+\omega))$ yields after a simple integral transformation
\beq
C(\tau) = \frac{1}{2\pi} \sum_\alpha \int\limits_{-\infty}^{+\infty} \Gamma_\alpha(\omega) \left[1+n_\alpha(\omega)\right] e^{+\ii\omega\tau} d\omega\,,
\eeq
which enables to directly read off the Fourier transform $\gamma(\omega) = \sum_\alpha \Gamma_\alpha(\omega) \left[1+n_\alpha(\omega)\right] \equiv \sum_\alpha \gamma_\alpha(\omega)$.
Evidently, the contributions of the two reservoirs enter additively in the rate equations, such that by labeling the energy eigenstates in the relevant
subspace ($a\to \f{n}$) we recover the rate equation with its coefficients stated in the main manuscript.

%%%%%%%%%%%%%%%%%%%%%%%%%%%%%%%%%%%%%%%%%%%%%%%%%%%%%%%%%%%%%%%%%%%%%%%%%%%%%%%%%%%%%%%%%%%%%%%%%%%%%%%%%%%%%%%%%%%%%%%%%%%
%%%%%%%%%%%%%%%%%%%%%%%%%%%%%%%%%%%%%%%%%%%%%%%%%%%%%%%%%%%%%%%%%%%%%%%%%%%%%%%%%%%%%%%%%%%%%%%%%%%%%%%%%%%%%%%%%%%%%%%%%%%

\subsection{Low Temperature Limit}

At sufficiently low temperatures, such that $n_\alpha(2 \epsilon_k) \ll 1$ for all $k\ge 3/2$ but still $n_\alpha(2 \epsilon_{1/2})=\ord(1)$, 
it is evident from Fig.~1 that most excited states will relax towards the two lowest states $\ket{0}\equiv\ket{0,0\ldots 0}$ and $\ket{1,0\ldots 0}\equiv\ket{1}$.
The dynamics in this subspace is governed by the rate equation
\beq
\left(\begin{array}{c}
\dot{\rho}_0\\
\dot{\rho}_1
\end{array}\right)
=
\abs{\bra{0} M_x \ket{1}}^2 \sum_\alpha \Gamma_\alpha(2\epsilon_{1/2}) 
\left(\begin{array}{cc}
-n_\alpha(2\epsilon_{1/2}) & +\left[1+n_\alpha(2\epsilon_{1/2})\right]\\
+n_\alpha(2\epsilon_{1/2}) & -\left[1+n_\alpha(2\epsilon_{1/2})\right]
\end{array}\right)
\left(\begin{array}{c}
{\rho}_0\\
{\rho}_1
\end{array}\right)
\eeq
with the matrix element
\beq
\abs{\bra{0} M_x \ket{1}}^2 = 4 u_{1/2}^2 v_{1/2}^2 
= \frac{s^2 \sin^2\left(\frac{\pi}{N}\right)}{1-2s(1-s)\left[1+\cos\left(\frac{\pi}{N}\right)\right]}\,,
\eeq
compare the discussion after Eq.~[5] in the main text.
Consequently, the current in this effective low-temperature limit becomes
\beq
I = \frac{\Gamma_S(2\epsilon_{1/2}) \Gamma_D(2\epsilon_{1/2}) \left[n_S(2\epsilon_{1/2})-n_D(2\epsilon_{1/2})\right]}
{\Gamma_S(2\epsilon_{1/2})\left[1+2 n_S(2\epsilon_{1/2})\right]+\Gamma_D(2\epsilon_{1/2})\left[1+2 n_D(2\epsilon_{1/2})\right]}
\frac{s^2 \sin^2\left(\frac{\pi}{N}\right)}{1-2s(1-s)\left[1+\cos\left(\frac{\pi}{N}\right)\right]}\,,
\eeq
which modifies the usual bosonic current through a two-level system by the matrix element $\abs{\bra{0} M_x \ket{1}}^2$.
Eventually, the $s$-dependence of this prefactor yields the non-monotonous dependence of the current on the phase-parameter $s$
in the low-temperature curves in Fig.~2.
%%%%%%%%%%%%%%%%%%%%%%%%%%%%%%%%%%%%%%%%%%%%%%%%%%%%%%%%%%%%%%%%%%%%%%%%%%%%%%%%%%%%%%%%%%%%%%%%%%%%%%%%%%%%%%%%%%%%%%%%%%%
%%%%%%%%%%%%%%%%%%%%%%%%%%%%%%%%%%%%%%%%%%%%%%%%%%%%%%%%%%%%%%%%%%%%%%%%%%%%%%%%%%%%%%%%%%%%%%%%%%%%%%%%%%%%%%%%%%%%%%%%%%%

\subsection{Non-equilibrium Stationary State}

The stationary solution of the rate equation can even for non-equilibrium (different temperature) configurations be obtained using basically two ingredients. 
First, we note that the Fourier transforms of the bath correlation functions obey the usual Kubo-Martin-Schwinger conditions
$\gamma_\alpha(-\omega) = e^{-\beta_\alpha \omega} \gamma_\alpha(+\omega)$, which lead when the system is coupled to only one
bath (e.g.\ by setting $\Gamma_D(\omega)=0$) to thermalization of the system with the temperature of the remaining reservoir (e.g.\ $\beta_S^{-1}$).
Formally, such a thermal state is characterized by the ratio of diagonal elements to be
\beq
\frac{\bar{\rho}_{\f{n}}}{\bar{\rho}_{\f{m}}} = e^{-\beta (E_{\f{n}}-E_{\f{m}})} = \frac{n(E_{\f{n}}-E_{\f{m}})}{1+n(E_{\f{n}}-E_{\f{m}})}\,,
\eeq
where $n(\omega)$ corresponds to the Bose distribution of the connected reservoir.
For coupling to multiple reservoirs we use that the occupations of the different reservoirs enter linearly and just weighted by the different tunneling rates to motivate
the ansatz ($\Delta_{\f{nm}} \equiv E_{\f{n}}-E_{\f{m}}$)
\beq
\frac{\bar{\rho}_{\f{n}}}{\bar{\rho}_{\f{m}}} = \frac{\bar{n}(\Delta_{\f{nm}})}{1+\bar{n}(\Delta_{\f{nm}})}\,,\qquad
\bar{n}(\omega) \equiv \frac{\Gamma_S(\omega) n_S(\omega)+\Gamma_D(\omega) n_D(\omega)}{\Gamma_S(\omega) + \Gamma_D(\omega)}\,.
\eeq
Indeed, one can easily prove for the rate equation
\beq\label{eq:rateeq}
\dot{\rho}_{\f{n}} &=& \sum_{\f{m}\neq \f{n}} \sum_\alpha \Gamma_\alpha (\Delta_{\f{mn}}) \left[1+n_\alpha(\Delta_{\f{mn}})\right] \abs{\bra{\f{n}} M_x \ket{\f{m}}}^2 \rho_{\f{m}}\nn
&&- \left(\sum_{\f{m}\neq \f{n}} \sum_\alpha \Gamma_\alpha(\Delta_{\f{nm}})\left[1+n_\alpha(\Delta_{\f{nm}})\right] \abs{\bra{\f{m}} M_x \ket{\f{n}}}^2 \right)\rho_{\f{n}}
\eeq
the validity of the stationary state by inserting 
\beq
\bar{\rho}_{\f{m}} = \frac{\bar{n}(\Delta_{\f{mn}})}{1+\bar{n}(\Delta_{\f{mn}})} \bar{\rho}_{\f{n}}
= \frac{\sum_\alpha \Gamma_\alpha(\Delta_{\f{mn}}) n_\alpha(\Delta_{\f{mn}})}{\sum_\alpha \Gamma_\alpha(\Delta_{\f{mn}}) \left[1+n_\alpha(\Delta_{\f{mn}})\right]} \bar{\rho}_{\f{n}}
\eeq
and using that $\Gamma_\alpha(\Delta_{\f{mn}}) = - \Gamma_\alpha(\Delta_{\f{nm}})$ and $n_\alpha(\Delta_{\f{mn}})=-\left[1+n_\alpha(\Delta_{\f{nm}})\right]$.
By a sequence of pair annihilations -- compare Fig.~1 in the main manuscript -- it therefore follows that any stationary occupation may be connected to 
the ground state occupation $\bar{\rho}_0$ via
\beq
\bar{\rho}_{\f{n}} = \bar{\rho}_0 \prod_{k>0} \left(\frac{\bar{n}(2\epsilon_k)}{1+\bar{n}(2\epsilon_k)}\right)^{n_k}\,.
\eeq
The latter is fixed by the normalization $\trace{\bar{\rho}_{\f{n}}}=1$
\beq
1 = \bar{\rho}_0 \sum_{n_{1/2}=0}^1 \ldots \sum_{n_{(N-1)/2}=0}^1 \prod_{k>0} \left(\frac{\bar{n}(2\epsilon_k)}{1+\bar{n}(2\epsilon_k)}\right)^{n_k}
= \bar{\rho}_0 \prod_{k>0} \left[\sum_{n_k=0}^1 \left(\frac{\bar{n}(2\epsilon_k)}{1+\bar{n}(2\epsilon_k)}\right)^{n_k}\right]
= \bar{\rho}_0 \prod_{k>0} \frac{1+2 \bar{n}(2\epsilon_k)}{1+\bar{n}(2\epsilon_k)}
\eeq
which yields~[7] in the manuscript.

%%%%%%%%%%%%%%%%%%%%%%%%%%%%%%%%%%%%%%%%%%%%%%%%%%%%%%%%%%%%%%%%%%%%%%%%%%%%%%%%%%%%%%%%%%%%%%%%%%%%%%%%%%%%%%%%%%%%%%%%%%%
%%%%%%%%%%%%%%%%%%%%%%%%%%%%%%%%%%%%%%%%%%%%%%%%%%%%%%%%%%%%%%%%%%%%%%%%%%%%%%%%%%%%%%%%%%%%%%%%%%%%%%%%%%%%%%%%%%%%%%%%%%%

\subsection{Stationary Energy}

Rewriting the system Hamiltonian~(\ref{Ehamplus}) as
\beq
H = \sum_{k>0} \epsilon_k \left(\gamma_k^\dagger \gamma_k + \gamma_{-k}^\dagger \gamma_{-k} -1\right)
\eeq
implies for its diagonal matrix elements in the relevant subspace $\bra{\f{n}} \HS \ket{\f{n}} = \sum_{k>0} \epsilon_k (2 n_k - 1)$.
The stationary expectation value of the system energy then becomes with~[7]
\beq
\expval{\bar{E}} &=& \trace{\HS \bar{\rho}} = \sum_{\f{n}} \bra{\f{n}} \HS \ket{\f{n}} \rho_{\f{n}} = \sum_{k>0} \epsilon_k \sum_{\f{n}} (2 n_k - 1) \rho_{\f{n}}\nn
&=& \sum_{k>0} \epsilon_k \sum_{n_k=0}^1 \frac{\left[\bar{n}(2\epsilon_k) \right]^{n_k}\left[1+\bar{n}(2\epsilon_k) \right]^{1-n_k}}{1+2 \bar{n}(2\epsilon_k)}(2 n_k-1)
= \sum_{k>0} \frac{-\epsilon_k}{1+2 \bar{n}(2\epsilon_k)}\,,
\eeq
where we have used that $\sum_{n_k=0}^1 \frac{\bar{n}^{n_k} \left[1+\bar{n}\right]^{1-n_k}}{1+2\bar{n}}=1$ holds for each $k$ separately in the second line.
In the thermodynamic limit ($N\to\infty$) and noting that all relevant quantities actually depend on $\kappa=k/N$, the
sum is easily converted into an integral, and we recover Eq.~[8] in the main text.

%%%%%%%%%%%%%%%%%%%%%%%%%%%%%%%%%%%%%%%%%%%%%%%%%%%%%%%%%%%%%%%%%%%%%%%%%%%%%%%%%%%%%%%%%%%%%%%%%%%%%%%%%%%%%%%%%%%%%%%%%%%
%%%%%%%%%%%%%%%%%%%%%%%%%%%%%%%%%%%%%%%%%%%%%%%%%%%%%%%%%%%%%%%%%%%%%%%%%%%%%%%%%%%%%%%%%%%%%%%%%%%%%%%%%%%%%%%%%%%%%%%%%%%

\subsection{Stationary Magnetization}

Similarly, we evaluate the diagonal matrix elements of the magnetization operator~(\ref{eq:mxreduced})
\beq
\bra{\f{n}} M_x \ket{\f{n}} &=& N - 4 \sum_{k>0} \abs{v_k}^2 - 4 \sum_{k>0} \left(\abs{u_k}^2-\abs{v_k}^2\right) n_k 
= N - 4 \sum_{k>0} \left[\abs{v_k}^2 + \left(1-2 \abs{v_k}^2\right) n_k\right]\,,
\eeq
which can be inserted in the stationary expectation value to yield
\beq
\expval{\bar{M}_x} &=& \sum_{\f{n}} \bra{\f{n}} M_x \ket{\f{n}} \bar{\rho}_{\f{n}} 
= N - 4 \sum_{k>0} \abs{v_k}^2 - 4 \sum_{k>0} \left(1-2\abs{v_k}^2\right) \sum_{n_k=0}^1 n_k 
\frac{\left[\bar{n}(2\epsilon_k) \right]^{n_k}\left[1+\bar{n}(2\epsilon_k) \right]^{1-n_k}}{1+2 \bar{n}(2\epsilon_k)}\nn
&=& N - 4 \sum_{k>0} \abs{v_k}^2 - 4 \sum_{k>0} \left(1-2\abs{v_k}^2\right) 
\frac{\bar{n}(2\epsilon_k)}{1+2 \bar{n}(2\epsilon_k)}
= N - 4 \sum_{k>0} \frac{\abs{v_k}^2 + \bar{n}(2\epsilon_k)}{1+2 \bar{n}(2\epsilon_k)}\,.
\eeq
Finally, the sum over $k$ can similarly be converted into an integral to yield~[9].
Furthermore, by inserting the coefficient~(\ref{eq:coefficients}) in the continuum representation and zero-temperature limit, 
we obtain for the magnetization density
\beq
\expval{\bar{m}_x} = \frac{\expval{\bar{M}_x}}{N} = 1 - 4 \int\limits_0^{1/2} v^2(\kappa) d\kappa 
= \frac{{\cal E}_E(4s(1-s))+(1-2s){\cal E}_K(4s(1-s))}{\pi(1-s)}\,,
\eeq
where ${\cal E}_E(x)$ and ${\cal E}_K(x)$ denote the complete elliptic integral and the complete elliptic integral of the first kind, respectively.

%%%%%%%%%%%%%%%%%%%%%%%%%%%%%%%%%%%%%%%%%%%%%%%%%%%%%%%%%%%%%%%%%%%%%%%%%%%%%%%%%%%%%%%%%%%%%%%%%%%%%%%%%%%%%%%%%%%%%%%%%%%
%%%%%%%%%%%%%%%%%%%%%%%%%%%%%%%%%%%%%%%%%%%%%%%%%%%%%%%%%%%%%%%%%%%%%%%%%%%%%%%%%%%%%%%%%%%%%%%%%%%%%%%%%%%%%%%%%%%%%%%%%%%

\subsection{Stationary Current}

The stationary current of bosons emitted to the drain can for example be obtained by inserting energy counting fields in the off-diagonal matrix elements
of the rate equation matrix, i.e., to perform in Eq.~(\ref{eq:rateeq}) the replacements
\beq
\Gamma_D(\Delta_{\f{mn}}) \left[1+n_D(\Delta_{\f{mn}})\right] &\to&  \Gamma_D(\Delta_{\f{mn}}) \left[1+n_D(\Delta_{\f{mn}})\right] e^{+\ii \Delta_{\f{mn}} \chi}\,,
\eeq
which automatically takes into account that $\Delta_{\f{mn}}>0$ corresponds to emission into the drain and $\Delta_{\f{mn}}<0$ to absorption.
Note that in the latter case one would use $\Gamma_D(-x) \left[1+n_D(-x)\right] = \Gamma_D(+x) n_D(+x)$.
This upgrades the rate equation by a counting field $\dot{\rho} = {\cal L}(\chi) \rho$, and the 
stationary current can then be obtained from the stationary state by deriving the rate matrix with respect to the counting field $\chi$
\beq
I &=& (-\ii) \trace{{\cal L}'(0) \bar{\rho}} 
= \sum_{\f{n}} \sum_{\f{m}\neq\f{n}} \Delta_{\f{mn}} \Gamma_D(\Delta_{\f{mn}})\left[1+n_D(\Delta_{\f{mn}})\right] \abs{\bra{\f{n}} M_x \ket{\f{m}}}^2 \bar{\rho}_{\f{m}}\nn
&=&  \sum_{\f{nm}\;:\;\Delta_{\f{mn}}>0} \Delta_{\f{mn}} \Gamma_D(\Delta_{\f{mn}})\left[1+n_D(\Delta_{\f{mn}})\right] \abs{\bra{\f{n}} M_x \ket{\f{m}}}^2 \bar{\rho}_{\f{m}}\nn
&&-\sum_{\f{nm}\;:\;\Delta_{\f{nm}}>0} \Delta_{\f{nm}} \Gamma_D(\Delta_{\f{nm}}) n_D(\Delta_{\f{nm}}) \abs{\bra{\f{n}} M_x \ket{\f{m}}}^2 \bar{\rho}_{\f{m}}\nn
&=& \sum_{\f{m}} \sum_{k>0} \left[2\epsilon_k m_k \Gamma_D(2\epsilon_k) \left[1+n_D(2\epsilon_k)\right] \left(4 u_k v_k\right)^2 \bar{\rho}_{\f{m}}
-2\epsilon_k (1-m_k) \Gamma_D(2\epsilon_k) n_D(2\epsilon_k) \left(4 u_k v_k\right)^2 \bar{\rho}_{\f{m}}\right]\nn
&=&\sum_{k>0} 2\epsilon_k \Gamma_D(2\epsilon_k) (4 u_k v_k)^2 \sum_{\f{m}}\left[m_k \left[1+n_D(2\epsilon_k)\right] - (1-m_k) n_D(2\epsilon_k)\right] \bar{\rho}_{\f{m}}\nn
&=&\sum_{k>0} 2\epsilon_k \Gamma_D(2\epsilon_k) (4 u_k v_k)^2 \sum_{m_k=0}^1 \left[m_k \left[1+n_D(2\epsilon_k)\right] - (1-m_k) n_D(2\epsilon_k)\right] 
\frac{\left[\bar{n}(2\epsilon_k) \right]^{m_k}\left[1+\bar{n}(2\epsilon_k) \right]^{1-m_k}}{1+2 \bar{n}(2\epsilon_k)}\nn
&=&\sum_{k>0} 2\epsilon_k \Gamma_D(2\epsilon_k) (4 u_k v_k)^2 \frac{\bar{n}(2\epsilon_k)-n_D(2\epsilon_k)}{1+2 \bar{n}(2\epsilon_k)}\nn
&=& \sum_{k>0} 2\epsilon_k (4 u_k v_k)^2 \frac{\Gamma_S(2\epsilon_k) \Gamma_D(2\epsilon_k) \left[n_S(2\epsilon_k) - n_D(2\epsilon_k)\right]}
{\Gamma_S(2\epsilon_k)\left[1+2 n_S(2\epsilon_k)\right]+\Gamma_D(2\epsilon_k)\left[1+2 n_D(2\epsilon_k)\right]}\,,
\eeq
which with evaluating the prefactor $A_k \equiv 4 u_k v_k$ from~(\ref{eq:coefficients}) becomes Eq.~[10] 
in the main manuscript.

The continuum representation of the current becomes (in wide-band limit $\Gamma_\alpha \equiv \Gamma_\alpha(2 \epsilon_k)$)
\beq\label{Ecurcont}
\frac{I}{N} &=&  32 \int\limits_0^{1/2} \frac{s^2 \Omega^2 \sin^2(2\pi\kappa)}{\epsilon(\kappa)}
\frac{\Gamma_S \Gamma_D [n_S(2\epsilon(\kappa))-n_D(2\epsilon(\kappa))]}{\Gamma_S[1+2n_S(2\epsilon(\kappa))]+\Gamma_D[1+2n_D(2\epsilon(\kappa))]} d\kappa
\equiv \int\limits_0^{1/2} j(s,\kappa) d\kappa\,.
\eeq
At the critical point and for small $\kappa$, the integrand behaves as
\beq
j(1/2,\kappa) &=& \frac{8 \pi \Omega (\beta_D-\beta_S) \Gamma_D \Gamma_S}{\Gamma_S \beta_D+\Gamma_D \beta_S} \kappa + \ord\{\kappa^2\}\,,\nn
\left.\frac{\partial}{\partial s} j(s,\kappa)\right|_{s=1/2} &=& \frac{32\pi\Omega(\beta_D-\beta_S) \Gamma_D \Gamma_S}{\Gamma_S \beta_D+\Gamma_D \beta_S} \kappa + \ord\{\kappa^2\}\,,
\eeq
which together with Eq.~[11] in the main manuscript leads to divergence of the second derivative of the current at the critical point for all temperature configurations.

This can also be seen in closed form in the infinite thermobias regime ($n_S(2\epsilon(\kappa))\to\infty$ and $n_D(2\epsilon(\kappa))\to 0$), where~(\ref{Ecurcont}) becomes
\beq
\frac{I}{N} &\to&
16 \Gamma_D (s \Omega)^2 \int\limits_0^{1/2} \frac{\sin^2(2\pi\kappa)}{\epsilon(\kappa)} d\kappa\nn
&=& \frac{4 \Gamma_D \Omega}{3\pi(1-s)^2} \left[(1-2s(1-s)) {\cal E}_E(4s(1-s)) - (1-2s)^2 {\cal E}_K(4s(1-s))\right]\,,
\eeq
where ${\cal E}_E(x)$ represents the complete elliptic integral and ${\cal E}_K(x)$ the complete elliptic integral of the first kind.

\end{document}